# LuminLab: An AI-Powered Building Retrofit and Energy Modelling Platform

KEVIN CREDIT,[1] QIAN XIAO,[2] JACK LEHANE,[1] JUAN VAZQUEZ,[2] DAN LIU,[2] LEO DE FIGUEIREDO[2]

[1]Maynooth University, Maynooth, Ireland
[2]Trinity College Dublin, Dublin, Ireland

ABSTRACT: *This paper describes the technical and conceptual development of the 'LuminLab' platform, an online tool that integrates a purpose-fit human-centric AI chatbot and predictive energy model into a streamlined front-end that can rapidly produce and discuss building retrofit plans in natural language. The platform provides users with the ability to engage with a range of possible retrofit pathways tailored to their individual budget and building needs on-demand. Given the complicated and costly nature of building retrofit projects, which rely on a variety of stakeholder groups with differing goals and incentives, we feel that AI-powered tools such as this have the potential to pragmatically de-silo knowledge, improve communication, and empower individual homeowners to undertake incremental retrofit projects that might not happen otherwise.*

KEYWORDS: *Energy efficiency, Building retrofit, AI, Machine learning, Pragmatism*

## 1. INTRODUCTION

Energy use in buildings represents a significant proportion of both household expenses and carbon emissions. European electricity prices have been steadily rising over the past 15 years, hitting an all-time high of €28.4 per 100 kWh in 2022 [1]. According to the International Energy Agency (IEA), operational building use amounts to almost 30% of global energy consumption [2]. To reach the IEA's 2030 net zero emissions target, buildings' operational energy use must be reduced by about 25%, requiring significant retrofit investment for existing building stock.

Unfortunately, the pursuit of net zero emissions targets and operational energy use in Ireland, and globally, is often restricted by the pressure of practical considerations and the misalignment of incentives and behaviours in policy creation. The expected benefit of retrofit packages for policy-makers often does not match with householders' thermal comfort demands [3] or the widespread availability (and cost) of conventional materials [4].

In addition, the building retrofit process in Ireland is complicated and costly for individual homeowners. As Figure 1(A) shows, required knowledge is often siloed in stakeholder groups and, despite the perceived linearity of retrofit processes diagrammed, phased service provision can further compound information incompatibility between stakeholders, further fragmenting information flow across time [5].

In most cases, even basic energy assessments can entail specialised design professionals and in-person energy audits that can cost in the range of €600-€800. Should a homeowner decide to proceed, they can spend significant time and effort pursuing individual quotes, planning and managing the project, not to mention temporary loss of use of their home.

Crucially, the options design professionals and home contractors offer often focus on deep retrofit solutions that may lack significant investment return for individual homeowners, even in the current high energy cost environment. Amid the availability of a broad range of technologies and major efforts to promote and accelerate retrofit activities, identifying the most effective retrofit strategies that meet investment criteria remains a significant challenge for homeowners [6]. There is very little information provided on smaller-scale solutions and simple fixes that could result in energy efficiency increases [7].

The development of new artificial intelligence (AI) technologies, such as large language models (LLM) like ChatGPT, provides a significant opportunity to improve this building retrofit process. By informing retrofit decisions across domains, AI can potentially help identify effective retrofit strategies [8], which could also support the shared creation of statistical models and cost-benefit estimation [9] for energy and economic efficiency [10] alike.

However, given the very recent development of many of these AI technologies, from a theoretical perspective there has been relatively little work contextualising the use of AI for building retrofit in general or understanding the extent to which AI-driven decision support systems can improve energy efficiency and stakeholder engagement in home retrofit projects. Here we conceptualise AI's use for building retrofit as a pragmatic solution to the communication and knowledge barriers that arise in these projects. As posited by Hillier [11], in order "to understand building, then, we must understand it both as a product and as a process". Pragmatism, as a knowledge claim based on "practical problem solving and real-world research" [12] offers considerable utility in addressing such measures as it is not committed to any specific school of thought, enabling researchers and designers to adopt the methods that best improve outcomes. Extending from this open and pluralistic epistemology, we envision AI tools as a part of a pluralistic methodology that can be used to help overcome the gaps in knowledge for individual homeowners, as well as the qualitative complexities of coordinating between stakeholder groups involved in retrofit projects. This offers particular significance in the multifaceted context within which energy retrofit research is nested, allowing for integration of practical action with the physical properties of retrofit packages — emphasising interconnectedness of buildings, systems, technologies and, critically, the stakeholder diversity required in retrofit solutions' design, production and eventual sustainment.

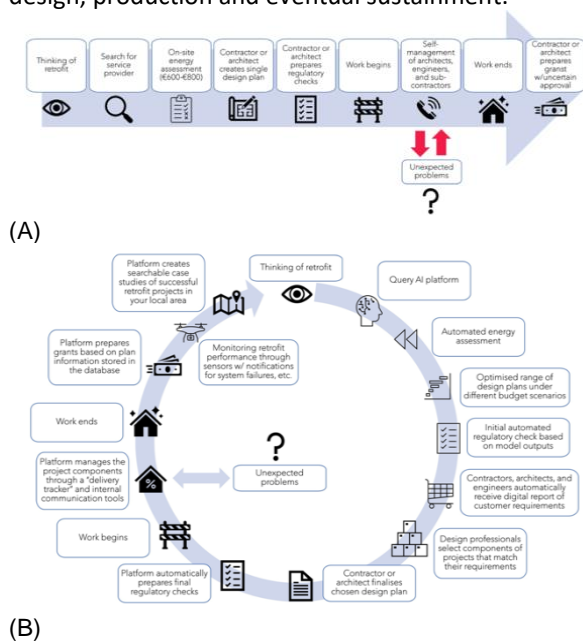

*Figure 1(A)-(B): Current vs. proposed homeowner retrofit management process.*

Given this pragmatic conceptual underpinning, the purpose of this paper is to describe the technical development of the 'LuminLab' AI-powered building retrofit platform created by the authors. This online tool integrates a purpose-fit human-centric AI chatbot and predictive energy model into a streamlined front-end that can rapidly produce and discuss building retrofit plans in natural language. As shown in Figure 1(B), the focus is on streamlining the end-to-end retrofit process by recognising its inherently recursive nature and eventually incorporating AI communication, modelling, and coordination tools at every stage, from project conception through planning, design, contractor assignment, coordination, and even monitoring. Providing an 'intelligent' chatbot agent that the user can interrogate helps to de-silo stakeholder knowledge and enable users to better understand the cost trade-offs inherent in the retrofit process.

Currently (as described below) the platform engages users with a range of possible retrofit pathways tailored to their individual building needs on-demand.

We finish the paper with reflections on the process and our ongoing engagements with relevant stakeholders in Ireland. While the tool is not yet widely implemented, our pragmatic, stakeholder-driven development approach provides useful lessons and a roadmap for future AI-driven support systems in the design professions within Ireland, and beyond.

## 2. METHODS

The LuminLab platform is an online application to streamline the home retrofit process, primarily for homeowners in Ireland. In Section 2.1 we describe the platform's front-end design and its integration with the LLM chatbot and back-end databases and processing. Then, in Section 2.2 we describe specifics of the energy model used to optimise retrofit configurations under a range of budget conditions.

### 2.1 Platform and LLM

At its core, the LuminLab platform uses a fine-tuned LLM as an intelligent interface to offer contextualised guidance for the entire retrofit process. The application is designed with a user-friendly front-end; upon accessing the platform, users encounter a dashboard which encompasses four primary components:

1. **Chat Assistant:** the LLM interface is the cornerstone of the LuminLab application, which acts as a central gateway to its various features and functionalities. Its integration with advanced models and technologies positions it as a dynamic and intelligent assistant, guiding users through every aspect of their home retrofitting journey.
2. **Reports and Plans:** this section aggregates all documents and plans generated during the user-chatbot interaction. It acts as an organised repository of platform-generated user-specific retrofit plans, analytical reports of the analysis of user-provided data (e.g. from their home's BER certificate), and tailored recommendations from the Chatbot based on respective user questions.
3. **My City in 3D:** this module (under development) allows users to explore their home's urban context in a dynamic 3D visualisation. In the future it will integrate geospatial data with LLM-aided analysis, presenting crucial information in an immersive and informative manner. Other planned future extensions include 3D modelling software integration to create detailed renderings of individual buildings using only homemade videos of the premises, a novel approach to significantly enhance visualisation and support the planning process. The underlying technology based on Neuralangelo, developed by NVIDIA, can be tailored to architecture modelling applications.

4. **Useful Resources:** to allow users to explore the information underpinning the LLM's responses, the platform also provides an extensive library of resources, including regulatory frameworks, retrofitting guidelines, and environmental data. This data is actively referenced by the LLM.

The front-end interface, crafted using React and Material-UI (MUI), is specifically designed to cater to a diverse user base. MUI's versatile component library allows rapid iteration of interface designs, ensuring an intuitive and accessible user experience. This adaptability is crucial in addressing users' varied technological comfort levels. To accommodate different user preferences, the application offers two modes of interaction: a structured multiple-choice form and a text-based chatbot. The multiple-choice form provides a clear, guided experience, asking specific questions with defined options for ease of data entry. The chatbot allows users to input data conversationally, capturing detailed retrofit information, also complementing data from the form. Users can utilise both modes concurrently, ensuring a comprehensive and tailored data submission process for various technological comfort levels. The specifics and methodologies employed in the multiple-choice form are further elaborated in Section 2.2 below.

The back-end of the platform, implemented using Flask, handles data processing and API interactions. The integration of the LLM, using the Langchain library, with vector databases hosted on Pinecone, enables access to a wide range of information sources which can include generated reports, government retrofit and building regulations, as well as previous conversations. At the time of writing, all development and testing has been carried out using the GPT-3.5 model, while also exploring open-source alternatives such as Llama2 13B and Mistral 7B. This facilitates the delivery of personalised and accurate retrofitting advice via a single unified communication system.

The chatbot is enhanced with access to an energy modelling machine learning model, capable of leveraging user-provided information to generate personalised home retrofit plans. Once the energy model generates the plans, they are initially structured JSON files, comprised of key-value pairs representing distinct plan components. To ensure compatibility and seamless integration with the LLM, these JSON files are transformed into a text-based format, where additional context is provided for each feature. This modified plan is incorporated into the LLM knowledge base, utilising a methodology consistent with the integration of user messages and additional input data. Following this process, the model is equipped to interpret these plans and provide the user with detailed explanations and responses to their queries. The LLM reads and interprets these plans, responding to users' queries with detailed explanations. This seamless interaction between the deep learning model and LLM allows for a comprehensive and user-friendly advisory system.

To improve user interaction and response nuances, a secondary LLM instance is trained on a large database of questions and corresponding follow-up queries, employing a vector similarity search approach to generate contextually appropriate and relevant follow-up questions shown in Figure 2.

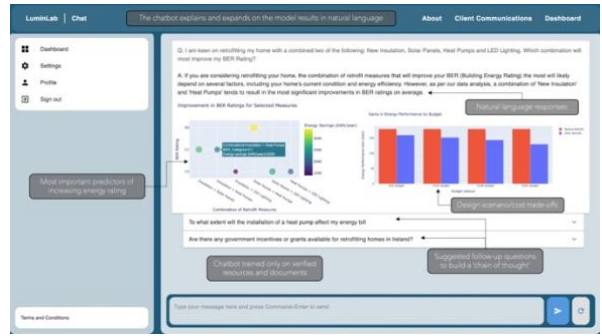

*Figure 2: Chatbot interface showing an example of an initial query, data visualisations, and follow-up questions.*

This feature is crucial in facilitating more meaningful interactions with users, ensuring their queries are comprehensively addressed by providing conversation paths they may not be aware of.

**2.2 Energy Model**

The energy model is used to help select appropriate retrofit components, generate detailed retrofit plans, and predict resulting energy ratings using deep learning methods. To do this, we use a classification model to predict the building features most related to increasing the building's energy performance rating, along with data on associated prices. This allows the user to select retrofit packages that correspond to their budget requirements.

The classification model is trained on the Energy Performance Certificates (EPC) dataset from the Sustainable Energy Authority of Ireland (SEAI) [13]. This contains over 1 million entries, categorised into 15 classes that correspond to different energy rating levels ('A1' being the best and 'G' being the worst). To build the predictive model we form 80%/10%/10% train/validation/test splits, where a different split is generated for each trial and all methods use the same splits. From the original 211-feature set that describe the individual building, we selected 41 features with both engineering and data-driven methods [14,15], ensuring the most relevant chosen features for our study. These 41 features are categorised into distinct groups based on different building aspects, including building envelope features like area of wall, roof, and door; building fabric features such as U-values for wall, roof, and door; heating system characteristics, notably main heating system efficiency; hot water-related attributes, including water storage volume; and spatial features such as the county code [14].

This project employs four deep learning algorithms to devise the classification model, namely multi-layer perceptron (MLP), self-supervised contrastive learning using random feature corruption (SCARF) [16], and coarse-to-fine-grained versions of MLP and SCARF. The MLP classifier is designed to determine the energy rating from selected building features. As Figure 3 shows, it consists of four hidden layers, with the input layer handling 41 feature dimensions. The data is passed through these hidden layers, with the class label predicted at output layer.

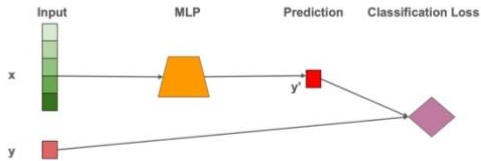

*Figure 3: Diagram showing multi-layer perceptron (MLP).*

We use the cross-entropy loss function to measure and minimise the difference between predicted and actual labels, aiding in the model's weight optimization. The subsequent stage involves fine-tuning, similar to the MLP classification process but with the initial phase's encoder translating input data into representations before MLP processing.

Meanwhile, SCARF operates in two stages, as shown in Figure 4. The initial unsupervised stage involves altering a randomly selected and replaced portion of the features — 30% in our case — based on a distribution derived from the training data [16]. Both the original and modified inputs are processed through an MLP-based encoder, generating respective representations. Here, the InfoNCE loss function is used, ensuring similarity between both corrupted and original inputs' representations [17].

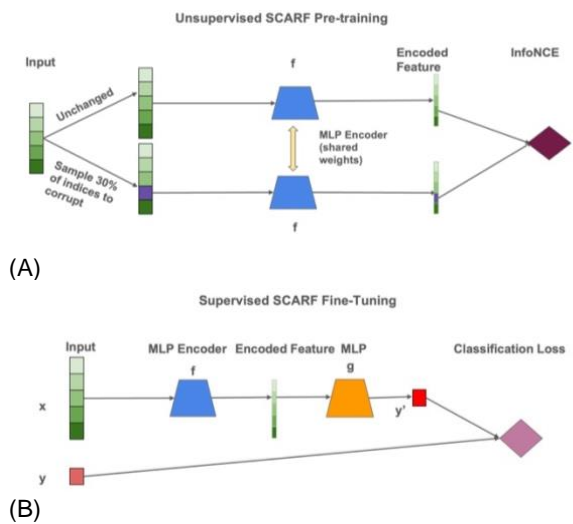

(A)

(B)

*Figure 4(A)-(B): Diagram showing unsupervised SCARF pre-training (A) and subsequent supervised finetuning (B). During pre-training, the encoder f is trained to learn a good representation for the input data. After pre-training, classification head g is applied on the top of f, and both g and f are subsequently fine-tuned for classification.*

The complexity in distinguishing between, e.g., classes A1, A2, and A3 in the EPC data led us to adopt a coarse-to-fine approach. Initially, we consolidated several classes into broader categories, reducing the original 15 classes to 5. This process involved two stages: first, training a coarse-grained classifier on the merged classes, followed by the development of five fine-grained classifiers for the subdivided classes, as depicted in Table 1. These coarse categories were then used in both the MLP and SCARF models.

To assess the highest performing method for use in the platform, we compare the MLP and SCARF deep learning approaches against three other popular machine learning algorithms used for classification problems, namely decision tree, gradient boosted tree and random forest. The performance results for each model are given in Table 2 (for macro F1 score and accuracy), demonstrating deep learning methods' outperformance of traditional machine learning methods when applied to the EPC dataset. MLP achieves the highest accuracy overall.

However, it is important to note that MLP shows limited accuracy in predicting certain classes, such as A1 in Table 3. This discrepancy can be attributed to the highly imbalanced nature of the EPC dataset, as there are very few A1 observations. In such datasets, accuracy may not be the most effective metric for assessing the model's performance on minority classes. Interestingly, SCARF and coarse-to-fine-grained classification significantly enhanced the model's performance on challenging classes like A1.

*Table 1: Mapping of original energy ratings (A1, A2, ..., G) to coarse categories (A, B, C, CD, EFG).*

| Original | A1 | A2 | A3 | B1 | B2 | B3 | C1 | C2 | C3 | D1 | D2 | E1 | E2 | F | G |
|---|---|---|---|---|---|---|---|---|---|---|---|---|---|---|---|
| Transformed | A | A | A | B | B | B | C | C | CD | CD | CD | EFG | EFG | EFG | EFG |

Despite these advancements, our models' overall performance is still not optimal. This limitation is largely due to the presence of missing fields and noise within the EPC dataset. For instance, we observed anomalies in the data, such as 10% of the floor area and floor U-values being recorded as zero. These missing observations are problematic, especially for U-values, where zero is not a plausible measurement.

*Table 2: Macro F1 scores and accuracies on the EPC test set. All results are averaged over 5 runs with different seeds.*

|  | Macro F1 | Accuracy |
|---|---|---|
| Decision Tree | 51.7% | 51.2% |
| Gradient Boosted Tree | 47.1% | 49.5% |
| Random Forest | 63.1% | 62.8% |
| MLP | 63.9% | 69.5% |
| SCARF | 65.8% | 68.7% |
| coarse-to-fine grained MLP | 66.8% | 68.3% |
| coarse-to-fine grained SCARF | 67.1% | 68.6% |

*Table 3: Accuracies on the EPC test set for class A1, A2 and A3. All results are averaged over 5 runs with different seeds.*

|  | A1 | A2 | A3 |
|---|---|---|---|
| MLP | 0% | 90.6% | 78.4% |
| SCARF | 17.5% | 86.1% | 83.4% |
| coarse-to-fine grained MLP | 36.8% | 90.2% | 79.6% |
| coarse-to-fine grained SCARF | 29.6% | 89.7% | 82.1% |

Substantively, our analysis revealed that building insulation features, particularly the U-values of walls and floors, have the most impact on these models. Note that model explanation tools such as LIME [18] and SHAP [19] are less reliable and effectively applied to deep learning models compared to linear or tree-based models. Therefore, we used the decision tree model's results to determine feature importance.

On the front-end, the LuminLab platform connects to the energy model; enabling users to retrofit homes by choosing from four key components: insulation (encompassing wall, roof, floor, window, door, and attic), boiler heating controls for temperature regulation, Mechanical Ventilation with Heat Recovery (MVHR) systems for air circulation, and solar panels (varying in number and power output)[1].

Within each component category, a variety of items are available for selection, each with associated characteristics and costs. For instance, in the category of door insulation, users can choose a door with specific properties, such as one made of aluminium with a U-value of 1.7, offered at a price of €1099. The system assumes a default comprehensive retrofitting approach, encompassing all components from wall insulation to the installation of solar panels. This default setting, however, may not align with users' actual requirements or preferences. Therefore, we empower users to customise retrofit components based on their unique needs and circumstances.

For each combination of items within different retrofit components, the model calculates a predicted energy rating. Each item in the combination influences specific home features. For example, selecting a roof with a U-value of 1.3 will adjust the home's original roof U-value to this new figure. Consequently, the model reflects such changes in home features which recalculates the energy rating accordingly. In addition, each item combination's cost is calculated, summing prices of all items in the plan. For each distinct energy rating obtained, we retain the combination with the lowest cost. Consequently, this determines the most cost-effective retrofit plan corresponding to each potential new energy rating. Moreover, any applicable grants from SEAI are also shown, aggregating item price minus potential grants. Figure 5 shows this general energy modelling process.

[1] Note: User inputs to the energy model are restricted to the information collected in the BER assessment, because the energy model uses only those features in the BER dataset to predict the best combination of elements (under the budget constraint) to improve BER. This means the information provided by the user is limited in terms of the level of detail (e.g., door jamb, glazed openings, etc.) to what is collected by BER assessors. However, future extensions of this project seek to be able to collect and incorporate finer levels of detailed information on the retrofit/configuration characteristics of homes.

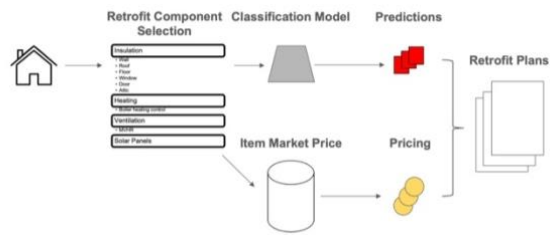

*Figure 5: Schematic of the energy modelling process.*

Ultimately, this system provides users with both energy rating post-retrofit and associated cost estimates for each proposed plan, enabling better-informed decisions for energy-efficient upgrades, particularly for fine-grained cost/efficiency trade-offs which are not always easy to estimate (even for knowledgeable homeowners) within the Irish market.

## 3. INITIAL FINDINGS AND CONCLUSIONS

While we have not yet widely tested the usage of the LuminLab platform in practice, initial findings and conclusions are presented from the development that may benefit retrofit scholars and practitioners.

First, from a global perspective: the platform's development – including everything from overall problem delineation to specific front-end design choices in the online tool – has been informed by conversations with a range of retrofit stakeholders in Ireland. These include government agencies and non-profit organisations such as the Sustainable Energy Authority of Ireland (SEAI), Dublin City Council (DCC), and Irish Green Buildings Council (IGBC), as well as individual homeowners, contractors, and architects.

Broadly, our stakeholder engagement attests to the lack of alignment between policy and practice. Given the technical nature of retrofit and decarbonisation policy, different groups perceive and engage with retrofit practice and measures of success differently. Despite each stakeholder group's perceived solutions to retrofit, each is often unaware of their disconnect with others' needs, priority areas, and lived experiences [20], attesting to the information fragmentation mentioned, compounded by the perceived linearity of the process in Figure 1(A) and 'phased' approach to service provision. For instance, policymakers create goals and standards that may have unintended consequences for contractors performing the work; homeowners may have alternate project goals (e.g., comfort) than what grants incentivise; and some specialised design knowledge may not be accessible in a small-scale project. This disconnection means that knowledge, resources, policies, and actions are siloed within stakeholder groups, making it difficult for individual homeowners to make sense of the complicated landscape that any retrofit project must traverse.

As for the specific technical components of the platform itself, we have found that enhancing LLMs with government data and regulations, carefully refined from web and document sources, is key to ensuring accurate and relevant responses in regulatory contexts. In addition, creating user-friendly default prompts for LLM-integrated websites is crucial, considering that most users might not be familiar with advanced prompting techniques. This approach aims to optimise tool performance and effectiveness for typical user interactions. For the energy model, we found that data irregularities from provided 'real-world' sources highlight the need for robust data pre-processing and model development to accommodate or correct data inconsistencies. While in this context the coarse-to-fine-grained SCARF model performs best, future work may focus on improving data quality and exploring models that are more resilient to data imperfections, thereby enhancing the reliability and applicability of our system in commercial settings. Substantively, our analysis also shows that the U-values of walls and floors have the most significant impact on improving the energy performance rating of a given building

Collectively, these learnings offer a critical path for de-siloing retrofit and AI's role in achieving this. Moreover, the pragmatist framework's conceptual distinctions underscoring our approach could better position further research inquiry, freeing researchers and participants to engage in a pluralism of approaches and methods to best serve the research question; the same pluralism underscoring the misalignment between technology and policy, and qualitative complexities of stakeholders involved. If properly channelled, an AI-informed and sufficiently multi-stakeholder-engaged retrofit can offer an alternative to extant policy and practice. This would enable reciprocal learning not otherwise possible for stakeholder groups individually for extended retrofit sustainment at both national and international levels.